\documentclass[11pt]{article}
\usepackage{amssymb}
\usepackage{amsmath}
\usepackage{fullpage}

\newcounter{rot}

\newcommand{\ignore}[1]{}

\def\a{\alpha} \def\b{\beta}  
   \def\g{\gamma}
 
 \def\th{\theta}   \def\l{\lambda}

 \def\om{\omega}

\newtheorem{lemma}{Lemma}
\newtheorem{theorem}{Theorem}

\newcommand{\proofstart}{{\bf Proof\hspace{2em}}}
\newcommand{\proofend}{\hspace*{\fill}\mbox{$\Box$}}

\newcommand{\ul}[1]{\mbox{\boldmath$#1$}}

\newcommand{\bfrac}[2]{\left(\frac{#1}{#2}\right)}

\newcommand{\brac}[1]{\left(#1\right)}

\newcommand{\set}[1]{\left\{#1\right\}}

\newcommand{\rai}{\rightarrow \infty}

\def\seq{\subseteq}

\newcommand{\ooi}{(1+o(1))}

\def\Pr{\mbox{{\bf Pr}}}

\newcommand{\gei}{\ge i}

\begin{document}

\begin{titlepage}
\title{ Chains-into-Bins Processes}
\author{Tu\u{g}kan Batu\thanks{Department of Mathematics, London
   School of Economics, London WC2A 2AE, UK. Email: {\tt
     t.batu@lse.ac.uk}.}
\and
Petra Berenbrink\thanks{School of Computing Science, Simon Fraser University,
 Burnaby, BC V5A 1S6, Canada. Email: {\tt petra@cs.sfu.ca}.}
\and
Colin Cooper\thanks{Department of  Computer Science, King's College
 London,  London WC2R 2LS, UK. Email: {\tt colin.cooper@kcl.ac.uk}.}}

\maketitle

\begin{abstract}
 The study of {\em balls-into-bins processes} or {\em occupancy problems} has a
 long history. These processes can be used to translate realistic problems
 into mathematical ones in a natural way.  In general, the goal of a
 balls-into-bins process is to allocate a set of independent objects (tasks,
 jobs, balls) to a set of resources (servers, bins, urns) and, thereby, to
 minimize the maximum load.  In this paper, we
 analyze the maximum load for the {\em chains-into-bins}
 problem, which is defined as follows. There are  $n$ bins, and $m$ objects to be allocated.
 Each object consists of balls connected into a
 chain of length $\ell$, so that there are $m \ell$ balls in total. We assume the chains cannot be broken, and that
 the balls in one chain
 have to be allocated to $\ell$ consecutive bins. We allow each chain
 $d$ independent and uniformly random bin choices for its starting position.
 The chain is
 allocated using the rule that the maximum load of any bin receiving
 a ball of that chain is minimized.  We show that, for $d \ge 2$ and
 $m\cdot\ell=O(n)$, the maximum load is $((\ln \ln m)/\ln d) +O(1)$
 with probability $1-\tilde O(1/m^{d-1})$.
\end{abstract}
\thispagestyle{empty}
\end{titlepage}

\section{Introduction}

The study of {\em balls-into-bins processes} or {\em occupancy
  problems} has a long history. These models are commonly used to
derive results in probability theory. Furthermore, balls-into-bins
processes can be used as a means of translating realistic
load-balancing problems into mathematical ones in a natural way.  In
general, the goal of a balls-into-bins process is to allocate a set of
independent objects (tasks, jobs, balls) to a set of resources
(servers, bins, urns).  It is assumed that the balls are independent
and do not know anything about the other balls.  Each ball is allowed
to choose a subset of the bins independently and uniformly at random
(i.u.r.) in order to be allocated into one of these bins.  The
performance of these processes is usually analyzed in terms of the
maximum load of any bin.

One extreme solution is to allow each ball to communicate with every
bin.  Thus, it is possible to query the load of every bin and to place
the ball into the bin that is least loaded.  This allocation process
always yields an optimum allocation of the balls. However, the time
and the number of communications needed to allocate the balls are
extremely large.  The opposite approach is to allow every ball to
communicate with only one bin. The usual model is for every ball to be
thrown into one bin chosen independently and uniformly at random.  For
the case of $m$ balls allocated to $n$ bins i.u.r., it is well known
that a bin that receives $m/n+\Theta\left( \sqrt{(m \log n)/n}\right)$
balls exists with high probability (w.h.p.).\footnote{A sequence of
  events $A_n$ occurs with high probability if $\lim_{n \rai} P(A_n)=
  1$.}  An alternative approach which lies between these two extremes,
is to allow every ball to select one of $d\ge 2$ i.u.r.\ chosen bins.
The GREEDY[$d$] process, studied by Azar et al.~\cite{ABKU99}, chooses
$d$ i.u.r.\ bins per ball, and the ball is allocated into the least
loaded among these bins. For this process, and $m \ge n$, the maximum
number of balls found in any bin, i.e., the {\em maximum load}, is
$m/n +\ln\ln n/\ln d +O(1)$ w.h.p.\ (see, for
example,~\cite{ABKU99},~\cite{BCSV06}).  Thus, even a small amount of
additional random choice can decrease the maximum load drastically
compared to a single choice.  This phenomenon is often referred to as
the ``power of two random choices'' (see~\cite{MRS00}).

In this paper we consider the chains-into-bins problem which can be regarded as a
generalization of the balls-into-bins problem.  We are given
$m$ chains consisting of $\ell$ balls each.  The balls of any chain have to
be allocated to $\ell$ consecutive bins. We allow each chain $d$ i.u.r.\ bin choices,
and allocate the chain using the rule that the maximum load of any bin
receiving a ball of that chain is minimized.
In this paper, we show that, for $d \ge 2$ and $m\cdot\ell =o(n\cdot
(\ln\ln m)^{1/2})$, the maximum load achieved by this algorithm is at
most $(\ln \ln m/\ln d)+O((m\ell/n)^2)+O(1)$, with probability
$1-O((\ln m)^d/m^{d-1})$.  This result shows that for a fixed number
of balls, the maximum load decreases with increasing chain length.
The maximum load depends on
the number of chains only in the following sense.  Allocating
$m=n/\ell$ chains of length $\ell$ (with a total number of $n$ balls)
into $n$ bins will, w.h.p., result in a maximum load of at most $\ln\ln
(n/\ell)/\ln d+O(1)$.
It
follows that if $\ell=O((\ln n)^a)$ for any $a>0$, our result is
asymptotically the same as that for allocating $n/\ell$ balls into
$n/\ell$ bins using GREEDY[$d$] protocol of~\cite{ABKU99}.

We also prove that the naive heuristic that `allocates the
chain headers using GREEDY[$d$] and hopes for the best' performs badly
for some values of $m,\ell$ as one might expect. Indeed, if $m \ge
\ln^2 n$ and $\ell\ge (\ln m)/\ln\ln m$, then the maximum occupancy of
this heuristic is at least $(\ln m)/(2\ln\ln m)$, w.h.p.

\paragraph{Clustering.}
It can be seen that, provided we can make some extra assumptions, the
results of~\cite{ABKU99} can be applied to the chains-into-bins
problem for $m$ chains of length $\ell$. Suppose we are {\em allowed
 to cluster the bins} into $N=n/\ell$ clusters of $\ell$ successive
bins, and each chain {\em can be allocated directly to one cluster}.
This is now equivalent to allocating $m$ balls into $N$ bins. Thus, we
would get a maximum load of $\Theta((\ln\ln N) /\ln(d)+1)$ with
GREEDY[$d]$ and $(\ln\ln N)/d+O(1)$ using the ALWAYS-GO-LEFT[$d$]
protocol.  However, this solution essentially ignores the model
under consideration, and is equivalent in a hashing context
to saying that we do not need to hash the data item at
the given location, but rather somewhere in the next $\ell$ cells at our convenience.
If we have this freedom to ignore the locations we
are given and pack the balls into $N=n/\ell$ clusters of length
$\ell$, then why not $N=n/(2\ell)$ clusters of length $2 \ell$,
placing the chains one after the other? Indeed why not arbitrary $N$,
or even $N=1$ and pack the chains cyclically in a round-robin fashion?
That would be even more efficient.  Obviously if such clustering were
available it would be easier to organize the behavior of the balls.
We assume henceforth that we have to put the chains where we are
instructed, rather than where we would like to. Finally, we remark
that, provided $m=N=n/\ell$, we can get the same results without
restructuring the problem, and thus, the extra provisions are
unnecessary.

\paragraph{Applications.}
Our model can be viewed as a
form of hashing in which the first data item of the chain is placed in the
selected position of the hash table, and the remaining items overflow into the
neighbouring positions of the table.

The chains-into-bins problem has several important applications.  One
example is data storage on disk arrays, such as RAID systems
(see~\cite{PGK88}).  Here, each data item is stored on several
neighbouring disks in order to increase the data transfer rate. In
this case, the bins model the disks from the storage array, and the
chains model data requests which are directed to several neighbouring
``bins.'' A second application is the scheduling of reconfigurable
embedded platforms (see~\cite{FKT01, SWP04}). Here, the tasks and the
reconfigurable chip are modelled as rectangles with integral
dimensions.  All tasks have the same height but different length. The
chip is modelled by a much larger rectangle that can hold several tasks
in both dimensions.  The goal is to allocate the tasks to a chip with
a fixed length such that the required height is minimized.  In this
case, the tasks are modelled by the chains and the chip is modelled by
the bins. The problem also models a scheduling problem where $m$ allocated
items persist in the system for $\ell$ time steps. For example,
 imagine a train travelling in a circle with $n$ station stops. The bins represent
stations and the length of a chain represent the number of stops travelled by
a passenger.

\subsection{Related Work}
Azar et al.~\cite{ABKU99} introduced GREEDY[$d$] to allocate $n$ balls into
$n$ bins. GREEDY[$d$] chooses $d$ bins i.u.r.\ for each ball and allocates the
ball into a bin with the minimum load.  They show that after placing $n$
balls, the maximum load is $\Theta((\ln\ln n) /\ln d)$, w.h.p. Compared to
single-choice processes, this is an {\em exponential} decrease in the maximum
load.  For the case where $m<n$, their results can be extended to show a
maximum load of at most $(\ln\ln n-\ln\ln (n/m))/\ln d +O(1)$, w.h.p.
V\"ocking~\cite{Voc03} introduced the ALWAYS-GO-LEFT[$d$] protocol, which
clusters the bins into $d$ clusters of $n/d$ consecutive bins each. Every ball
now chooses i.u.r.\ one bin from every cluster and is allocated into a bin with
the minimum load. If several of the chosen bins have the same minimum load,
the ball is allocated into the ``leftmost'' bin.  The protocol yields a
maximum load of $(\ln\ln n)/d+O(1)$, w.h.p.  In~\cite{KP06}, Kenthapadi and
Panigrahy suggest an alternative protocol yielding the same maximum load. They
cluster the bins into $2n/d$ clusters of $d/2$ consecutive bins each. Every
ball now randomly chooses $2$ of these clusters and it is allocated into the
cluster with the smallest total load. In the chosen cluster, the ball is then
allocated into the bin with minimum load again. The authors also argue in that
paper that clustering is essential to reduce the load to $(\ln\ln n)/d+O(1)$.
In~\cite{BCSV06}, the authors analyse GREEDY[$d$] for $m\gg n$.  It is shown
that the maximum load is $m/n+\ln\ln (n)$, w.h.p.  Mitzenmacher et
al.~\cite{mps03} show that a similar performance gain occurs if the process is
allowed to memorize a constant number of bins with small load.

In~\cite{SV03}, Sanders and V\"ocking consider the {\em random arc allocation
 problem}, which is closely related to the chains-into-bins problem. In their
model, they allocate arcs of an arbitrary length to a cycle.  Every arc is
assigned a position i.u.r.\ on the cycle. The chains-into-bins problem with
$d=1$ can be regarded as a special discrete case of their problem, where the
cycle represents the $n$ bins and the arcs represent the chains
(in~\cite{SV03}, different arc lengths are allowed). Translated into the
chains-into-bins setting, the authors show the following result.  If
$m=n/\ell$ chains of length $\ell$ are allocated to $n$ bins ($m \rai$), then
the maximum load is at most $(\ln (n/\ell))/(\ln\ln (n/\ell))$, w.h.p.  Note
that their result is asymptotically the same as that for allocating $n/\ell$
balls into $n/\ell$ bins, provided that $n/\ell\rightarrow\infty$.
In~\cite{matt}, the author shows that the expected maximum load is smaller if
we allocate $n/2$ chains of length $2$ with one random choice
per chain, compared to $n$ balls into $n$ bins with $d=2$.


\section{Model and Results}\label{chains}

Assume $m$ chains of length $\ell$ are allocated i.u.r.\ to bins wrapped
cyclically round $1,...,n$. A chain contains $\ell$ balls linked together
sequentially.  The first ball of a chain is called {\em header}; the remaining
balls comprise the {\em tail} of the chain.  If chain $i$ (meaning the header
of chain $i$) is allocated to bin $j$, then the balls of the chain occupy
bins $j, j+1,\ldots, j+\ell-1$, where counting is modulo $n$.  We define
the {\em h-load} of a bin as the number of headers allocated to the bin.  This
is to be distinguished from the {\em load} of bin $j$.  The load
is the total number of balls allocated to bin $j$; that is, the number of chain
headers allocated to bins $j-\ell+1,\ldots, j-1,j$.

We consider the case where each chain header randomly chooses $d$ bins
$j_1,\ldots, j_d$. For random choice $j_k$ it computes the maximum load of
bins $j_k,j_{k}+1,\ldots j_{k}+\ell-1$.  The chain header is allocated to the
bin $j_k\in j_1,\ldots, j_d$ such that the maximum load is minimized. This
allocation process is called GREEDY\_CHAINS[$d,\ell$].

We show the following result, which is proved in Section~\ref{agc}.

\begin{theorem}\label{dchain}
Let $m \le n$, $\ell \ge 1$, $d\ge 2$, and assume $m\cdot\ell\ge n/(2e)$ and
$m\cdot \ell =o(n(\ln\ln m)^{1/2})$. Let $m$ chains of length $\ell$ be
allocated to $n$ bins with $d$ i.u.r.\ bin choices per chain header.
The maximum load of any bin obtained by GREEDY\_CHAINS[$d,\ell$] is at
most
\[
\frac{\ln \ln m}{\ln d}+O\left(\left(\frac{m\cdot\ell}{n}\right)^2\right),
\]
with  probability $1-O((\ln m)^d/m^{d-1})$.
\end{theorem}

Note that when $m\cdot \ell =O(n)$, GREEDY\_CHAINS[$d,\ell$] achieves a
maximum load of $(\ln \ln m)/(\ln d) + O(1)$, with high probability.
In order to make a direct comparison, we extend of the results of~\cite{ABKU99}
on the algorithm GREEDY[$d$] to the case where $m<n$.

\begin{theorem}\label{1achain}
 Assume that $d \ge 2$, $m\le n$, and that $c$ is an arbitrary
 constant. Then, the maximum load achieved by GREEDY[$d$] after the
 allocation of $m$ balls is at most
\[
\frac{\ln \ln n-\ln\ln(n/m)}{\ln d } +c
\]
with a probability of $1-O(n^{-s})$, where $s$ is a constant depending
on $c$.
\end{theorem}

Theorem~\ref{1achain} gives the bin load arising from chain headers (ignoring
the rest of the chain). Since other collisions can occur, for example, between
chain headers and internal links of the chain, this will always be a lower
bound on the maximum load.  Then, provided $m \ell/n=O(1)$,
\[
\frac{\ln \ln n-\ln \ln (n/m)}{\ln d} +O(1) \le
\; \mbox{max load} \;
\le \frac{\ln \ln m}{\ln
 d}+O(1).
\]
We see that, provided that $\ell=e^{(\ln n)^{o(1)}}$ (in particular,
$\ell$ is poly-logarithmic in $n$), the ratio of the upper and lower
bounds on the maximum load is $\ooi$.

Finally, suppose we allocate chain headers 
using GREEDY[$d$] but ignore the effect of this allocation on the rest
of the chain.  The following theorem, proved in Section~\ref{ahead},
shows that this approach leads to a large maximum occupancy.

\begin{theorem}\label{nogood}
Assume that $m\cdot\ell \ge n/(2e)$, that $m< n/(2e)$, that $m\ge \ln^2 n$, and that
$\ell \ge (\ln m)/(\ln \ln m)$. Then, the maximum occupancy of any bin
based on GREEDY[$d$] allocation of chain headers is at least $(\ln m)/(2\ln \ln m)$,
w.h.p.
\end{theorem}

The proofs of Theorem~\ref{1achain} and Theorem~\ref{nogood} can be
found in Section~\ref{ahead}.

\section{Analysis of GREEDY\_CHAINS[$d,\ell$] }\label{agc}
In this section, we prove Theorem~\ref{dchain}.  The proof uses
layered induction.  In the case of GREEDY[$d$], Azar et
al.~\cite{ABKU99} use variables $\gamma_i$ as a high-probability upper
bound on the number of bins with $i$ or more balls, where
$\gamma_6={n}/{2e}$ and, for $i>6$,
$\gamma_i= e\cdot n\cdot(\gamma_{i-1}/n)^{d}$.

Since we allocate chains into bins, we cannot consider only the number
of bins with $i$ or more chain headers, we have to consider both the
chain headers and tails.  Hence, to calculate the load of a bin, we
have to consider the chain headers allocated to neighbouring bins. To
do so, we define the set $S_i$ which can be thought of as the set of
bins which will result in a maximum load of at least $i+1$ if one of
the bins in $S_i$ is chosen for a chain header. The set $S_i$ contains
all bins $j$ with load (at least) $i$ and the bins at distance at most
$\ell-1$ in front of bin $j$.  We emphasize that not all bins in $S_i$
have load of $i$ themselves. We use variables $\b_i$ as
high-probability upper bounds and show that, for $i$ large enough,
$|S_i|\leq \b_i =2e\cdot m\cdot \ell\cdot (\b_{i-1}/n)^d$, w.h.p., in
our induction. In the following, we define some sets and random
variables that are used in our analysis.

\begin{itemize}
\item Let $\l_j(t)$ be a random variable counting the {\em h-load} of bin $j$.
 That is, $\l_j(t)$ is the number of chain headers allocated to bin $j$ at
 (the end of) step $t$ for $t=0,1,...,m$.

\item For given $A \seq [n]$, define $\l_A(t)=\sum_{j \in A} \l_j(t)$.  Thus,
 $\l_A(t)$ is a random variable counting the total {\em h-load} of the bins
 in $A$ at the end of step $t$.

\item Let $R_j=\set{j-\ell+1,...,j-1,j}$, the set of bins that will increase
 the load of bin $j$ if a chain header is allocated to them.

\item Let $L_j(t)$ be a random variable counting the {\em load} of bin $j$ at
 (the end of) step $t$.  Thus, $L_j(t)=\l_{R(j)}(t)$, the load arising from
 the chain headers allocated to the $\ell$ bins of $R(j)$.

\item Let $Q_i(t)=\set{j:L_j(t) \ge i}$ be the set of labels of bins
whose load is at least $i$ at (the end of) step $t$.

\item Let $S_i(t)= \cup_{j \in Q_i(t)} R_j$.  Thus, $S_i(t)$ contains the
 labels of bins such that an allocation of a chain header to one of
 these bins will increase the load of a bin with a load of at least
 $i$ by 1.

\item Let $\th_{\gei}(t)=|S_i(t)|$.

\item Let $h_t$ be a random variable counting the height of chain $t$.  The
 algorithm GREEDY\_CHAINS[$d, \ell$] allocates the header
 to the bin which minimizes the maximum total load $h_t$, where
\begin{equation}\label{ht}
h_t=1+ \min_{i=1,...,d}\;\;\max \;\set{L_{j_i+k}(t-1),\;k=0,...,\ell-1}
\end{equation}
and $j_1,...,j_d$ are the bins chosen i.u.r.\ at step $t$.
\end{itemize}

Our method of proving Theorem~\ref{dchain} uses an approach developed
in~\cite{ABKU99}, but incorporates the added complexity of considering
the maximum load over the chain length.  For consistency, we have
preserved notation as far as possible.

Let $\a=m\ell/ n$ with $\a\ge 1/2e$ and $k=\lceil8\alpha^2 e\rceil$.
First, we show that $i$ chains can contribute a block of bins
of length at most $2\ell-1$ to $S_i(m)$.

\begin{lemma}\label{extcase}
 For $i\ge k$, (1) $\th_{\gei}(m)\le 2m\ell/i$, (2) $\th_{\gei}(m)
 \le n/2$.
\end{lemma}
\proofstart To prove part~(1) we first consider a the following worst case
scenario. Suppose at step $t$ bin $j$ contains $i$ chain headers, bins
$j-\ell+1,...,j-1$ are empty, and bins $j+1,...j+\ell-1$ do not contain any
chain headers. Then, $\set{j,j+1,...j+\ell-1}\seq Q_i(t)$,
$\set{j-\ell+1,...,j,j+1,...j+\ell-1}\seq S_i(t)$ and $|S_i(t)|\ge 2\ell-1$.

Now suppose that $|S_i(t)|=r$ and ask the question how many
chain headers do we need for that. This number is minimized when the
chain headers are aligned, as demonstrated in our worst case
example. Then, every set of $i$ chain headers covers $2\ell-1$
bins. This means that we need
$$r\ge \frac{i \cdot |S_i(t)|}{2\ell-1}$$
chain headers. In general, for $t\le m$,  we get
\[
\l_{S_i(t)}(t) \ge
\frac{i \cdot |S_i(t)|}{2\ell-1} > \frac{i\cdot \th_{\gei}(t)}{2\ell}
\]
since $\th_{\gei}(t)=|S_i(t)|$.
For $t=m$, we have
\begin{equation}\label{L1}
\frac{i \cdot \th_{\gei}(m)}{2\ell}\le \l_{S_i(m)}(m)\le m.
\end{equation}

Part~(2) follows from part~(1).
Since  $i\ge k$, $\alpha= m\ell/n$, and $\alpha\ge 1/(2e)$, we get
$$\th_{\ge i}(m)\le\th_{\ge k}(m)\le \frac{2m\ell}{k}\le\frac{2m\ell}{8\alpha^2 e}=
\frac{2m\ell}{8\cdot(m\ell/n)\cdot\alpha e}= \frac{n}{4\alpha e}\le
\frac{n}{2}.$$
\proofend

To prove Theorem~\ref{dchain}, we define
\[
\b_i=\left\{ \begin{array}{ll}
n&i=1,...,k-1;\cr
\frac{n}{4e\a}& i=k;\cr
2e m\ell \cdot\bfrac{\b_{i-1}}{n}^d & i > k.
\end{array}\right.
\]
For $i \ge 0$ and $j=k+i$, it follows from the definition of $\beta_i$ that
\begin{equation}
\label{ubd}
\b_{j}=\b_{k+i}=
n \cdot \frac{(2e)^{\frac{d^i-1}{d-1}}}{(4e\a)^{d^i}}=
n\cdot 2^{-d^i}\cdot (2e\a)^{-(d^i(d-2)+1)/(d-1)},
\end{equation}
and, thus, provided $2e\a\ge 1$ (i.e., $m\ell/n\ge 1/2e$), we have
$\b_{k+i}\le n\cdot 2^{-d^i}$.

Define
$
{\mathcal E}_i(t)=\set{\th_{\gei}(t)\le \b_i}$
and let
\begin{equation}\label{e0}
{\mathcal E}_i={\mathcal E}_i(m)=\set{\th_{\gei}(m)\le \b_i}
\end{equation}
be the event that $|S_i(t)|$ is bounded by $\b_i$ throughout the
process. From the discussion following~\eqref{L1}, we have that
${\mathcal E}_{k}$ holds with certainty. Our goal is to obtain a value
for $i$ such that $\Pr({\mathcal E}_i)$ is close to 1 and, given
${\mathcal E}_i$, no bin receives more than $i$ balls, with high
probability.

We next state a standard lemma (see Lemma 3.1 in~\cite{ABKU99}), proof
of which is omitted.

\begin{lemma} \label{Lem3.1}
Let $X_1,X_2,...,X_m$ be a sequence of random variables with
values in an arbitrary domain, and let $Y_1,Y_2,...,Y_m$
be a sequence of binary random variables with the property that
$Y_t=Y_t(X_1,...,X_t)$. If
\[
\Pr(Y_t=1 \mid X_1,...,X_{t-1}) \le p,
\]
then
\[
\Pr\brac{\sum_{t=1}^m Y_t \ge k} \le \Pr(B(m,p)\ge k),
\]
where $B(m,p)$  denotes a binomially distributed
random variable with parameters $m$ and $p$.
\end{lemma}

As the $d$ choices of bins for a chain header are independent, we have that
\[
\Pr\brac{h_t \ge i+1 \mid \th_{\gei}(t-1)=r} \le \left(\frac{r}{n}\right)^d,
\]
where $h_t$ is given by Eq.~\eqref{ht}. For chain $t$ and integer $i$,
let $Y_t^{(i)}$ be an indicator variable given by
\[
Y_t^{(i)}=1 \iff \set{h_t \ge i+1, \quad \th_{\gei}(t-1) \le \b_i}.
\]

Let $X_i=(x_i^1,\ldots, x_i^d)$ denote the set bin choices of $i$th
chain header, and let $X_{1,t}=(X_1,...,X_{t})$ be the choices of the
first $t$ chains.  We define $\ul X_{1,t}$ as the event $\{\ul
X_{1,t}=(X_1,...,X_{t})\}$.

Assume ${\ul X_{1,t-1}} \in {\mathcal E}_i(t-1)$,
meaning after the allocation of the first $t-1$ chains, we have at
most $\beta_i$ bins that would result in a load of at least $i+1$ when
hit by chain $t$.  Then, $$ \Pr(Y_t^{(i)}=1 \mid{\ul
 X_{1,t-1}})\le\bfrac{\b_i}{n}^d$$ and, if ${\ul X_{1,t-1}}\not \in
{\mathcal E}_i(t-1)$,
then $\Pr(Y_t^{(i)}=1 \mid{\ul  X_{1,t-1}})=0$. Either way,
\[
\Pr(Y_t^{(i)}=1 \mid{\ul X_{1,t-1}} )\le \bfrac{\b_i}{n}^d \stackrel{\Delta}{=} p_i.
\]
We can apply Lemma~\ref{Lem3.1}  to conclude that
\begin{equation}\label{L3}
\Pr(\sum_{t=1}^{m} Y_t^{(i)} \ge r) \le \Pr(B(m,p_i) \ge r).
\end{equation}


Considering the {extreme case} discussed above in
Lemma~\ref{extcase}, we see that each event $\{Y_t^{(i)}=1\}$ adds
at most an extra $2\ell -1 $ bins to $S_{i+1}(t)$.
Thus for $\ul  X_{1,m} \in {\mathcal E}_i$,
\begin{equation}\label{L4}
\th_{\ge i+1}(m) \le 2\ell  \cdot \sum_{t=1}^m Y_t^{(i)}.
\end{equation}
Let $r_i=e\cdot m\cdot p_i$. Then, provided that $\sum_{t=1}^m
Y_t^{(i)} \le r_i$, we have
\begin{equation}\label{e3}
 \th_{\ge i+1}(m) \le 2 \ell r_i = 2 \ell em\cdot p_i= 2e m\ell \cdot
 \bfrac{\b_i}{n}^d= \b_{i+1}.
\end{equation}

From~\eqref{L3} and~\eqref{L4}, we have
\begin{equation}\label{e1}
 \Pr\left(\th_{\ge i+1}(m) > 2\ell \cdot r_i \,\Bigm| {\mathcal E}_i\right) \le
 \Pr\left(\sum_{t=1}^m Y_t^{(i)} > r_i \,\Bigm| {\mathcal E}_i\right) \le
 \frac{\Pr(B(m,p_i)\ge r_i)}{\Pr({\mathcal E}_i)}.
\end{equation}
Provided that $m\cdot p_i \ge 2 \ln\om$ (where the precise value of $\om$
is established below in \eqref{omega}), using the Chernoff bounds, we get
\begin{equation}\label{e2}
\Pr(B(m,p_i) \ge em\cdot p_i)\le e^{-m\cdot p_i}\le\frac{1}{\om^2}.
\end{equation}

Recall that $\Pr(\neg {\mathcal E}_{k})=0$.  Assume inductively that
$\Pr(\neg{\mathcal E}_{i})\le i/\om^2$, for $i \ge k$.  Since
\[
\Pr(\neg {\mathcal E}_{i+1}) \le \Pr(\neg {\mathcal E}_{i+1}\mid
{\mathcal E}_{i})\cdot \Pr({\mathcal E}_{i})+\Pr(\neg{\mathcal E}_{i}),
\]
we have, from \eqref{e0}, \eqref{e3}, \eqref{e1}, and \eqref{e2}, that
\[
\Pr(\neg {\mathcal E}_{i+1}) \le \frac{i+1}{\om^2}.
\]

Choose $i^*$ as the smallest $i$
such that $p_i=\bfrac{\b_i}{n}^d \le \frac{2 \ln \om}{m}$.
From~\eqref{ubd},
\begin{equation}\label{boundistar}
i^*-k \le \frac{\ln\ln (m/\ln \om) }{\ln d} +O(1).
\end{equation}
Also, as $\a=m\ell/n=o((\ln\ln m)^{1/2})$, we have that $k=o(i^*)$ so
that the induction is not empty. Since $p_{i^*}\le(2\ln \om)/m$, we
have that
\begin{eqnarray*}
\Pr(\th_{\ge i^*+1}(m) \ge (2 \ell)\cdot 6 \ln \om \mid {\mathcal E}_{i^*})
&\le& \frac{\Pr(B(m,p_{i^*})\ge 6 \ln \om)}{\Pr({\mathcal E}_{i^*})}\\
&\le& \frac{\Pr(B(m,(2 \ln \om)/m)\ge 6 \ln \om)}{\Pr({\mathcal E}_{i^*})}\\
&\le& \frac{1}{\om^2\cdot \Pr({\mathcal E}_{i^*})},
\end{eqnarray*}
and, thus,
\begin{equation}
\label{bd}
\Pr(\th_{\ge i^*+1}(m) \ge (2 \ell) \cdot 6 \ln \om)\le
\frac{1}{\om^2\cdot \Pr({\mathcal E}_{i^*})}\cdot\Pr({\mathcal E}_{i^*})+
\Pr(\neg {\mathcal E}_{i^*})
\le (i^*+1)/\om^2.
\end{equation}


Conditioned on $\theta_{\ge i^*+1}(m) \le 12 \ell \ln \omega$, the
probability that a chain is placed at height at least $i^*+2$ is at
most $(12 \ell \ln \omega/n)^d$. Given that $Y \sim B(m,(12 \ell \ln
\omega/n)^d)$, $\Pr(Y\ge 1) \le m(12 \ell \ln \omega/n)^d$ by
Markov's Inequality.  Thus applying Lemma~\ref{Lem3.1}, we get

\begin{equation}\label{theotherbit}
\Pr\left(\sum_{t=1}^m Y_t^{(i^*+1)} \ge 1 \Bigm|  \th_{\ge i^*+1} (m)\le
 (2 \ell) \cdot  6\ln \om \right) \le
\frac{m \cdot \bfrac{12\ell\cdot \ln \om}{n}^d}{\Pr(\th_{\ge i^*+1}
\le 12\ell\cdot \ln \om)}.
\end{equation}
Let $ \om$ satisfy
\begin{equation}
\label{omega}
\om=\bfrac{m^{d-1}\cdot\ln\ln m}{(\ln m)^d}^{1/2}.
\end{equation}

Using $m\cdot \ell =o(n(\ln\ln
m)^{1/2})$,~\eqref{omega},~\eqref{theotherbit} and~\eqref{bd}, the
probability that there is a bin with load at least $i^*+2$ is
bounded by a term of order
\[
\frac{i^*+1}{\om^2}+ O\left(\frac{(\ln \om)^d}{m^{d-1}}\right) =
 O\bfrac{(\ln m)^d}{m^{d-1}}.
\]

By plugging in $\om$ and $k$ into~\eqref{boundistar}, we get
\[
i^* \le \frac{\ln\ln m}{\ln d} + O(\alpha^2) + O(1)
.\]
Thus, w.h.p, no bin receives more than $i^*+1$ balls. \proofend


\section{Allocating Chain Headers}\label{ahead}

In this section we present the proofs of Theorem~\ref{1achain} and Theorem~\ref{nogood}.

\subsection{Proof of Theorem~\ref{1achain}}
This theorem can be shown similarly to the proof of Theorem 4
in~\cite{ABKU99}. We define $\gamma_1=\gamma_2=\cdots=\gamma_5=n$,
$\gamma_6=m/(2e)$, and
$$\gamma_i=em\cdot \bfrac{ \gamma_{i-1}}{ n}^d \mbox{ for } i>6.$$
Thus $\g_i=C\;n(m/(n2e))^{d^i}$ for some $C>1$ constant.
Integer $i^*$ is defined as the smallest $i$ such that $em(\gamma_i/n)^d\le 6\ln n$,
which holds for
$$i^*\le \frac{\ln\ln n-\ln\ln (n/m)}{\ln d}+O(1)
.$$
It can be shown that the maximum load is bounded by $i^*+2=(\ln\ln
n-\ln\ln(n/m))/\ln d+O(1)$, w.h.p.

\subsection{Proof of Theorem~\ref{nogood}}

The idea
of the proof is as follows. Let $U_m$ be the number of bins with load
at least one after the allocation of $m$ chains by GREEDY[$d$] applied
to the chain headers. We show that, with a good probability, there
exists a strip of $\ell$ consecutive bins which i)~is used by one
chain, and ii)~at least $t$ of its bins are in $U_m$. This gives us a
bin with load at least $t$.

First, we find a lower bound on $U_m$.  Azar et al.~\cite{ABKU99} show
that the protocol GREEDY[$d$] for $d\ge 1$ is majorized by GREEDY[1] in the
following sense. Let $x_i$ be the load of the bin with the $i$th
largest load after allocation of $m$ balls with GREEDY[$d$], and let
$x'_i$ be the load of the bin with the $i$th largest load after
allocation of $m$ balls with GREEDY[$1$].  It is shown
in~\cite{ABKU99} that there exists a one-to-one mapping between the
random choices of GREEDY[$1$] and GREEDY[$d$] such that for all $1\le
j\le n$,
$$\sum_{i=1}^j x_i\le \sum_{i=1}^{j} x'_i.$$
From this it follows that the number of empty bins in an allocation
with GREEDY[$d$] is smaller that the one in GREEDY[$1$].  Let $f(m)$
be the number of occupied bins in an allocation generated by
GREEDY[$1$].  When $m=n/(2e)$ we get
$$E[f(m)]=n\brac{1-\brac{1-\frac{1}{n}}^{m}} \ge 0.9m.$$ As $m$ decreases,
$m-E[f(m)]$ decreases, too.
Thus, provided $m\le n/(2e)$, the expected number of occupied cells with
$d\ge 1$ choices per ball is always at least $0.9m$. By concentration, we can
assume $U_m \ge m/2 +1$, provided $m \ge \ln^2 n$.

Given $U_m$, we can assume that the locations of the bins occupied by
chain headers are sampled uniformly without replacement from
$[1,\dots,n]$.  We fix one of the $m$ chains and consider the strip
of~$\ell$ consecutive bins occupied by the chain. Assuming $\ell \ge
2t$, the probability of at least $t$ bins occupied by additional chain
headers in that strip is at least
\[{U_m-1 \choose t} \cdot (\ell)_t \cdot \bfrac{1}{n}^t \ge
{m/2 \choose t} \cdot (\ell)_t \cdot \bfrac{1}{n}^t \ge\bfrac{m\ell}{2e
 t n}^t \ge \bfrac{1}{4e^2t}^t,
\]
as $m\ell \ge n/(2e)$ and $(\ell)_t=\ell(\ell-1)\cdots(\ell-t+1) \ge
(\ell/e)^t$.

Let $c=1/(4e^2)$.
Then, the expected number of chains allocated into a strip with load
at least $t=\ln m/(2\ln\ln m)$
is at least
\begin{eqnarray*}
m\cdot \bfrac{c}{t}^t&=& \exp \brac{\ln m-t\ln t/c}\\
&\ge & m^{1/3},
\end{eqnarray*}
for large enough $m$. The probability that such an event does not
occur tends to zero by the Chebychev's inequality.\proofend


\section{Conclusions and Open Problems}

In this paper we analyse the maximum load for the {\em
  chains-into-bins} problem where $m\cdot\ell$ balls are connected in
$m$ chains of length $\ell$.  We show that, provided $m\ell\ge n/2e$
and $m\ell =o(n(\ln\ln m)^{1/2})$, the maximum load is at most
$\frac{\ln \ln m}{\ln d}+O((m\ell/n)^2)$, with probability $1-O((\ln
m)^d/m^{d-1})$.  This shows that the maximum load is going down with
increasing chain length.

Surprisingly, there are many open questions in the area of balls-into-bins
processes. Only very few results are known for weighted balls-into-bins processes,
where the balls come with weights and the load of a bin is the sum of the
weights of the balls allocated to it.  Here, it is even not known if two or
more random choices improve the maximum load, compared to the simple process
where every ball is allocated to a randomly chosen bin
(see~\cite{tw07}). Also, it would be interesting to get tight results for the
maximum load and results specifying ``worst-case'' weight distributions for
the balls. Something in the flavor ``given that the total weight of the balls
is fixed, it is better to allocate lots of small balls, compared to fewer
bigger ones.''  Another interesting problem is to show results relating the
maximum load to the order in which the balls are allocated. For example, is it
always better to allocate balls in the order of decreasing ball weight,
compared to the order of increasing ball weight?

For chains-into-bins problem, an open question is to prove
Knuth's~\cite{knuth} conjecture stating that breaking chains into two parts
only increases the maximum load. This question still open for a single choice
and also for several random choices per ball. See~\cite{matt} for a first
progress in this direction. Another question is if similar results to the one
we showed in this paper for GREEDY[$d$] applied to chains also holds for the
ALWAYS-GO-LEFT protocol from~\cite{Voc03} applied on chains.

Finally, we note that it would be interesting to generalize the
problem to two dimensional packing, and consider online allocation of
$m$ objects of length $\ell$ and width $w$ to the cells of a toroidal
grid of length $n$ and width $h$.

\bibliographystyle{plain}



\end{document}